\documentstyle{article}
\title{\sc BLACKBODY DISTRIBUTION FOR WORMHOLES}
\author{ Pedro F. Gonz\'alez-D\'{\i}az.\\
Instituto de Matem\'aticas y F\'{\i}sica Fundamental\\
Consejo Superior de Investigaciones Cientificas\\
Serrano 121, 28006 Madrid (SPAIN)\\
}
\date{December 30,1992}
\begin{document}
\maketitle
\large
\setlength{\baselineskip}{0.9cm}
\vspace{3cm}

Classification numbers: 0460;0490

\pagebreak

ABSTRACT

By assuming that only (i) bilocal vertex operators which are diagonal with
respect
to the basis for local field operators, and (ii) the convergent elements with
nonzero positive energy of the density matrix representing the quantum state
of multiply-connected wormholes, contribute the path integral that describes
the effects of wormholes on ordinary matter fields at low energy,
it is obtained that the probability measure for multiply-connected wormholes
with
nondegenerate energy spectrum is given in
terms of a Planckian probability distribution for the momenta of a quantum
field
$\frac{1}{2}\alpha_{l}^{2}$, where the $\alpha_{l}$'s are the Coleman
parameters,
rather than a classical gaussian distribution law, and that an observable
classical universe can exist if, and only if, such multiply-connected wormholes
are
allowed to occur

\pagebreak

\section{INTRODUCTION}

In this paper we shall re-explore some consequences arising from the spectrum
dependence
of the wormhole dynamics. In particular, we shall concentrate on the
effects that the insertion of multiply-connected wormholes [1,2] may have on
ordinary
matter fields at low energies in the asymptotic regions from a statistical
standpoint. It will be seen that the sums over the number of equal-state
wormholes
and over the different wormhole states, required to obtain the wormhole
distribution function, are not independent of each other. Thus, the present
study supersedes earlier work on nonsimply connected wormhole dynamics done [1]
without imposing any dependence between the two sums, and without explicitely
taking
into account the relative probability for the states. This relative
probability,
which moreover appears as a prefactor in the bi-local interaction expression
for
nonsimply connected wormholes [1], should quite naturally enter the path
integral
expressing the effects of wormholes on ordinary matter, just at the same
footing
as the semiclassical probability for wormholes does.

\section{PROBABILITY MEASURE FOR QUANTUM GRAVITY}

In the dilute wormhole approximation [3], the effects of a single wormhole on
ordinary
matter fields in the asymptotic regions can be expressed in the path integral
for the expectation value of a given observable $O$ by inserting a factor
which,
if the wormhole quantum state is given by a density matrix, is
$-\frac{1}{2}C_{ij}\beta^{2}\epsilon_{mn}^{-1}$.
Adapting the formalism given by Klebanov, Susskind and Banks [4] to this case,
we have then
\begin{equation}<O>  \propto \int
dgOe^{-I(g,\lambda)}(\frac{1}{2}\Sigma_{i,j}C_{ij}\beta_{i}(x)\beta_{j}(y)\epsilon_{mn}^{-1}),\end{equation}
where $C_{ij}\propto e^{-S_{w}}$, in which $S_{w}$ is the Euclidean action for
the wormhole,
$\epsilon_{mn}=E_{m}^{(f)}-E_{n}^{(g)}$, with the $E'$s being the energy
levels for the matter field ($f$) and gravitational field ($g$) harmonic
oscillators, respectively, and $\epsilon_{mn}^{-1}$ the relative probability
for the state $\Psi_{mn}$;
$\lambda$ collectively denotes parameters such as
coupling constants, particles masses, the cosmological constant, etc.;
$\beta_{i}=\frac{1}{2^{\frac{1}{2}}}\int d^{4}xg(x)^{\frac{1}{2}}K_{i}(x)$,
where $K_{i}$ denotes the vertex operator and
the index $i$ labels the elements of a basis for the local field operators at
the point $x$ on the large region. This interpretation ensures the grouping of
the $\alpha_{i}$ parameters with the coupling constants of a generic
Lagrangian [4]. Indices $i$,$j$ are independent of the quantum numbers $n$,$m$
which label the off shell energy spectrum [1]. All dependence of the path
integral
on that
spectrum is incorporated through the relative probability factor
$\epsilon_{mn}^{-1}$
because a unique invariant quantum theory of wormhole must satisfy the
boundary requirement that the possible different kinds of wormholes and
wormhole
states ought all to have an equal asymptotic (classical) behaviour, thus
rendering the insertion amplitude to join their ends onto the asymptotic
regions
independent of the wormhole spectrum.

It is worth noticing that, besides the
usual insertion amplitude depending only on the vertex function at each
insertion
point on the low-energy large regions which corresponds to wormhole pure
states,
for mixed states the wormhole effective interaction factor also depends on the
absolute value [1,5] of the eigenenergies $\epsilon_{mn}$. This additional
dependence arises because the wormhole is now off shell and has, therefore, a
true energy spectrum. Clearly, for Planck-size wormholes, which is the case
we shall consider throughout this paper, we have $\epsilon_{mn}=m-n$.

Eqn.(1) supersedes Eqn.(4.11) of Ref. 1 which was prepared for performing the
sum over the states first.

The multiply-connected wormhole dynamics derived from (1) will be re-considered
in what follows. According to Coleman [6], we first sum over any number of
wormholes, all in the state with energy $\epsilon_{mn}$. Then (1) exponentiates
to give
\begin{equation}\int dg O
e^{-I(g,\lambda)}e^{\frac{1}{2}\sum_{ij}C_{ij}\beta_{i}\beta_{j}(m-n)^{-1}}.\end{equation}
The exponent in (2) is still bi-local. However, it can be made local by using
the transformation employed by Coleman [6]. Eqn. (2) can then be re-written
\begin{equation}\int dg O
e^{-I(g,\lambda)}\int\prod_{p}d\alpha_{p}e^{-\frac{1}{2}(m-n)D_{ij}\alpha_{i}\alpha_{j}}e^{-\beta_{l}\alpha_{l}},\end{equation}
with
\begin{equation}D_{ij} = C_{ij}^{-1} = e^{S_{w}},\end{equation}
through which the position-independent parameter $\alpha$ enters the formalism.

Now, unlike the case considered by Coleman in which there was no true wormhole
spectrum, one should next sum over the quantum numbers $m$ and $n$; i.e. over
all possible wormhole states. However, since multiply-connected wormhole
physics
at the inner (Planck) region contributes only through the quantum-number
combination $m-n$ appearing in the second exponent of (3), in summing over
$m$ and $n$ independently, there will be an overcounting (in the sum over the
number of wormholes)
associated with all
those values of $m$ and $n$ that result in the same value of their difference
$m-n$. This overcounting, which would actually lead to summing again over any
number of wormholes,
can be however gauged off by simply replacing the independent
sums over $m$ and $n$ by a single sum over the discrete index defined by
$k=m-n$. Negative values of $k$ would be associated with divergent path
integrals [1].
However, such negative values of $k$ are here ruled out because, once the
path integral has been made local in (3), there could no longer be
gravitational
excitation energy equal to or greater than the matter excitation energy, and
therefore
$k>0$, since, otherwise, relative to an observer in one asymptotic region, the
resulting negative energy density could convert any decreasing cross-sectional
area into an increasing cross-sectional area near the wormhole throat, and
this would necessarily represent a bi-local insertion, in contradiction with
the local character of the path integral after applying the Coleman's
transformation
leading to (3). This assumption may also be connected with the feature [7] that
Euclidean nonsimply-connected wormholes describable as a nonfactorizable
density matrix are squeezed states which require violation of the weak energy
condition,
but simply connected wormholes do not. Making the path integral local somehow
renders the wormhole manifold divided into two disconnected parts, each now
describable as an unsqueezed pure state which should not violate weak energy
condition [8].
On the other hand, the lower limit in this summation should be taken
to be 1 rather than 0 because there always are wormhole instantons (of the
Tolman-Hawking type [3]) which are off shell even in the pure gravity
limit [9]. We have then
\[\sum_{k=1}^{\infty}\int dg O
e^{-I(g,\lambda)}\int\prod_{p}d\alpha_{p}e^{-\frac{1}{2}kD_{ij}\alpha_{i}\alpha_{j}}e^{-\alpha_{l}\beta_{l}}\]
\begin{equation}= \int\prod_{p}d\alpha_{p}\int dg O e^{-I(g,\lambda +
\alpha)}(e^{\frac{1}{2}D_{ij}\alpha_{i}\alpha_{j}}-1)^{-1},\end{equation}
where the third exponent $-\alpha_{l}\beta_{l}$ in the l.h.s. of (5) has been
inserted
in the path integral of the large universe in the r.h.s.. Hence, the path
integral can be written in the form
\begin{equation}<O> \propto \int d\alpha P(\alpha)Z(\alpha)<O>_{\lambda +
\alpha},\end{equation}
where
\begin{equation}P(\alpha) =
\frac{1}{e^{\frac{1}{2}D_{ij}\alpha_{i}\alpha_{j}}-1},\end{equation}

\begin{equation}Z(\alpha) = \int dg e^{-I(g,\lambda + \alpha)}.\end{equation}
The probability measure will be then
\begin{equation}\mu(\alpha) = P(\alpha)Z(\alpha).\end{equation}

The set of equations (6)-(9) is the main result of this work. It differs from
the similar result obtained by Coleman in that the probability distribution for
the effective coupling constants $\alpha$ is given here by Eqn. (7) which
reflects an essentially $quantum$ uncertainty about the values of $\alpha$:
instead of the $classical$ measure obtained for wormholes in a
pure quantum state, one obtains a probability distribution which closely
resembles the Planck formula for black-body radiation for the most general
mixed
wormhole states. In what follows, we shall use the shorthand notation
$D\alpha^{2}$
for $D_{ij}\alpha_{i}\alpha_{j}$. Actually, we will see in the next section
that
among all possible combinations of indices $i$,$j$ in (7), only the diagonal
combinations $i=j=l$ are allowed by quantum requirements, so that the Coleman
probability becomes gaussian and (7) reduces to
\begin{equation}
P(\alpha)\equiv P(\alpha_{l}) = \frac{1}{e^{\frac{1}{2}D\alpha_{l}^{2}}-1}.
\end{equation}

\section{DOES QUANTUM GRAVITY REQUIRE AN EXTRA QUANTIZATION?}

The classical gaussian probability law,
$P_{c}(\alpha)$, discovered by Coleman [4,6]
\begin{equation}P_{c}(\alpha) = e^{-\frac{1}{2}D\alpha^{2}}.\end{equation}
does not represent a true probability distribution, but
just an unrenormalized probability. However, one may still introduce a
normalized
probability distribution for Coleman statistics however by defining
\begin{equation}p_{c}(\alpha) =
\frac{e^{-\frac{1}{2}D\alpha^{2}}}{\sum_{\alpha}e^{-\frac{1}{2}D\alpha^{2}}}.\end{equation}
A true probability measure can now be derived from (12) by performing the
statistical average for the path integral $Z(\alpha)$, i.e.
\begin{equation}\bar{Z}(\alpha) =
\sum_{\alpha}Z(\alpha)p_{c}(\alpha).\end{equation}
Clearly, if we want to obtain $\mu(\alpha)$ as given by (9) and (7) from
(13), one must necessarily require that neither $\alpha$ nor $Z(\alpha)$ can
vary continuously, but are integral multiples of some minimum values
$\alpha_{0}$
and $Z_{0}\equiv Z(\alpha_{0})$, such that $\alpha^{2}\equiv
\alpha_{n}^{2}=n\alpha_{0}^{2}$,
and $Z(\alpha)\equiv Z_{n}=nZ_{0}$. In this case,
\begin{equation}\mu\equiv\bar{Z}(\alpha) =
\frac{Z(\alpha)\sum_{n=0}^{\infty}ne^{-nx}}{\sum_{n=0}^{\infty}e^{-nx}} =
\frac{Z(\alpha)}{e^{\frac{1}{2}D\alpha^{2}}-1} =
Z(\alpha)P(\alpha),\end{equation}
where for the sake of generality, we have dropped off the subscript $0$ for
$\alpha$,
$x=\frac{1}{2}D\alpha^{2}$, and $Z(\alpha)=\frac{1}{2}\alpha^{2}$. We shall
show
in what follows that the last identification is actually a necessary condition
for a consistent derivation of $Z(\alpha)P(\alpha)$ from (13).

Thus, starting with the Coleman probability for the $\alpha$ parameters, we
have recovered the main result of this work. It follows that the statistical
character of the most general quantum state for wormholes appears to lead to
an essential discontinuity which is over and above that is already contained in
Coleman theory.

For large $D\alpha^{2}$, we recover then the Coleman probability measure
$P_{c}(\alpha)Z(\alpha)$; in turn, for small $D\alpha^{2}$, one obtains the
semiclassical probability law, $\mu(\alpha)=D^{-1}$. Thus, if we interpret
$D^{-1}$
as an equilibrium temperature, and the $\frac{1}{2}\alpha^{2}$ as the momenta
of a quantum field, Coleman distribution and the probability for Euclidean
gravity (here, $e^{-S_{w}}$) appear to play the role of, respectively, the
Wien law and the Rayleigh-Jeans law for average energy. This interpretation
is made more precise if we calculate the quantity
\begin{equation}\omega=\int Dd\mu = \ln\frac{[\frac{2(\mu +
Z)}{\alpha^{2}}]^{\frac{2(\mu +
Z)}{\alpha^{2}}}}{[\frac{2\mu}{\alpha^{2}}]^{\frac{2\mu}{\alpha^{2}}}} +
Const.,\end{equation}
and interpret it as the entropy of the system when it contains just one large
universe. For $N$ large universes, we had then
$N\mu=P\nu$ (with $P$ an integer), and it is finally obtained
\begin{equation}\Omega=N\omega=\ln\frac{(P+N)^{P+N}}{N^{N}P^{P}} + Const.,
\end{equation}
where we have used $Z=\nu =\frac{1}{2}\alpha^{2}$. Now, if we interpret (16)
as an entropy, the probability measure (14) is immediately recovered from
just the condition [10] that $\Omega$ is a maximum, which corresponds to a
system in
thermal equilibrium at temperature $D^{-1}$. We can see now why
$Z=\frac{1}{2}\alpha^{2}$
is a necessary condition to achieve full consistency in this interpretation.
Furthermore, if $Z(\alpha)=\frac{1}{2}\alpha^{2}$, it also follows that only
diagonal elements $i=j=l$ may enter the quantity $\alpha_{i}\alpha_{j}$,
since according to (8) the path integral $Z(\alpha)$ must only depend on just
one index labelling the basis for the local field operators. Hence,
$Z(\alpha)\equiv Z(\alpha_{l})$
and the probability distribution reduces to (10).

In what follows we shall explore some possible implications that the existence
of multiply-connected wormholes may have in the convergence problem of
Euclidean
quantum gravity. It has been recently emphasized by Hawking [11] that the
divergence problem of Euclidean path integral becomes even more severe in the
presence
of wormholes. In this case, besides the known divergences arising from the
unboundedness of the Euclidean action, there appears a very serious  divergence
in the probability measure on the space of effective coupling constants which
is connected to the Coleman
mechanism for the vanishing of the cosmological constant. None of these
divergences should be expected to have any relevance if the wormholes are off
shell. The path integral $Z(\alpha)$ becomes then discontinuous as well, and
the relevant quantum state for the large universe with an arbitrary number of
multiply-connected wormholes is given by the probability measure itself, that
is, by the statistical average of the path integral $Z(\alpha)$. The point
now is to check that the so-defined quantum state is finite under all
circunstances.
That this is actually the case can be readily seen by performing the integral
\begin{equation}\Theta = \int_{0}^{\infty}dZ(\alpha)\mu(\alpha) =
\frac{\pi^{2}}{6}D^{-2},\end{equation}
which is in fact finite.

Convergence will be preserved even if we just integrate over $\alpha$ as the
Riemann zeta function $\zeta(z)$ is regular for all values of $z$ other
than unity [12]. Moreover, introducing the concept of probability measure
density,
\begin{equation}\rho_{\mu}(\alpha) = \mu(\alpha)Z(\alpha)^{2},\end{equation}
and integrating over $Z(\alpha)$, one obtains
\begin{equation}U_{\mu} = \int_{0}^{\infty}dZ(\alpha)\rho_{\mu}(\alpha) =
\frac{\pi^{4}}{15}D^{-4}.\end{equation}

Clearly, Eqn. (19) plays the role of a Stefan-Boltzmann law for quantum
gravity,
and states that the total probability measure density is given by the fourth
power of the nucleation rate of baby universes per Planck volume per Planck
time.
Actually, Eqns. (17) and (19) express the very remarkable result that an
observable universe can exist if, and only if, it is connected to a nonzero
number of multiply-connected wormholes. This result can quite naturally be
connected  with the recent proposal [2,7] of a universal decoherence in the
matter field sector by which all quantum correlations contained in the pure
state-vector evolving according to the Schroedinger equation, are traced off by
the action of multiply
connected wormholes.

In summary, this paper shows that, if we assume bilocal vertex operators
diagonal
with respect to the indices that label the elements of the basis for local
field
operators on the large regions, and consider that only convergent density
matrix
elements contribute the path integral for the quantum state of
multiply-connected
wormholes,
the effects of such multiply-connected wormholes with nondegenerate energy
spectrum
on ordinary matter at low energies are given in terms of a Planckian
probability
distribution law for the momenta of a quantum field
$\frac{1}{2}\alpha_{l}^{2}$.
Under these assumptions,
it has been also seen
that for such a distribution law to admit a consistent interpretation, it is
required that both, the $\alpha$ parameters themselves and the path integral
for
the large universes should vary discountinuously, and that this path integral
be also given by $\frac{1}{2}\alpha^{2}$. It appears, moreover, that all
divergences
of quantum gravity could be remedied according to this interpretation which
parallels closely Planck theory of black-body radiation.
\vspace{2cm}

$Acknowledgements$: This work was supported by an Accion Especial of C.S.I.C.
and a CAICYT Research Project N' PB91-0052.
The author wants to thank S.W. Hawking, P. Tinyakov and C.L. Siguenza for
useful discussions.
\pagebreak

REFERENCES

[1] P.F. Gonz\'alez-D\'{\i}az 1991 Nucl. Phys. B351 767

[2] P.F. Gonz\'alez-D\'{\i}az 1992 Phys. Rev. D45 499

[3] S.W. Hawking 1987 Phys. Lett. B195 337; 1988 Phys. Rev. D37 904

[4] I. Klebanov, L. Susskind and T. Banks 1989 Nucl. Phys. B317 665

[5] P.F. Gonz\'alez-D\'{\i}az, in: The Physical Origin of Time Asymmetry,
eds. W. Zurek, J.J. Halliwell, and J. P\'erez-Mercader
(Cambridge
Univ. Press, Cambridge, 1993), p. 413; "Topological Arrow of Time", Instituto
de
Optica Preprint QGC-06/91 (unpublished).

[6] S. Coleman 1988 Nucl. Phys. B310 643

[7] P.F. Gonz\'alez-D\'{\i}az, 1992 Phys. Lett. B293 294.

[8] D. Hochberg and T.W. Kephart 1991 Phys. Lett. B268 377

[9] P.F. Gonz\'alez-D\'{\i}az 1989 Phys. Rev. D40 4184

[10] T.S. Kuhn, Black-Body Theory and the Quantum Discontinuity, 1894-1912
(University Chicago Press, Chicago, 1978)

[11] S.W. Hawking 1990 Nucl. Phys. B335 155

[12] M. Abramowitz and I.A. Stegun, Handbook of Mathematical Functions (Dover,
New York, 1972)

\end{document}